\documentclass[acmtog]{acmart}

\definecolor{amethyst}{rgb}{0.6, 0.4, 0.8}
\definecolor{amaranth}{rgb}{0.9, 0.17, 0.31}
\definecolor{blueviolet}{rgb}{0.54, 0.17, 0.89}

\usepackage{enumitem}
\usepackage{array}
\usepackage{multirow}
\usepackage{soul}
\usepackage{hyperref}
\newcolumntype{P}[1]{>{\centering\arraybackslash}p{#1}}

\AtBeginDocument{%
  \providecommand\BibTeX{{%
    \normalfont B\kern-0.5em{\scshape i\kern-0.25em b}\kern-0.8em\TeX}}}

\usepackage{etoolbox}
\makeatletter
\patchcmd\@combinedblfloats{\box\@outputbox}{\unvbox\@outputbox}{}{%
}%

\acmSubmissionID{891}

\usepackage[capitalise,noabbrev,nameinlink]{cleveref}
\begin{document}

\setcopyright{acmlicensed}
\acmJournal{TOG}
\acmYear{2023} \acmVolume{42} \acmNumber{6} \acmArticle{225} \acmMonth{12} \acmPrice{15.00}\acmDOI{10.1145/3618399}

\title{BakedAvatar: Baking Neural Fields for Real-Time Head Avatar Synthesis}

\author{Hao-Bin Duan}
\email{duanhb@buaa.edu.cn}
\orcid{0009-0002-1200-3236}
\affiliation{%
  \institution{State Key Laboratory of Virtual Reality Technology and Systems, Beihang University}
  \city{Beijing}
  \country{China}
  \postcode{100191}
}

\author{Miao Wang}

\authornote{Corresponding Author: Miao Wang (miaow@buaa.edu.cn)}
\email{miaow@buaa.edu.cn}
\orcid{0000-0002-4102-8582}
\affiliation{%
  \institution{State Key Laboratory of Virtual Reality Technology and Systems, Beihang University, and Zhongguancun Laboratory}
  \city{Beijing}
  \country{China}
  \postcode{100191}
}

\author{Jin-Chuan Shi}
\email{jinchuanshi@buaa.edu.cn}
\orcid{0009-0003-4899-2205}
\affiliation{%
  \institution{State Key Laboratory of Virtual Reality Technology and Systems, Beihang University}
  \city{Beijing}
  \country{China}
  \postcode{100191}
}

\author{Xu-Chuan Chen}
\orcid{0009-0006-7072-4305}
\affiliation{%
\institution{State Key Laboratory of Virtual Reality Technology and Systems, Beihang University}
  \city{Beijing}
  \country{China}
}

\author{Yan-Pei Cao}
\email{caoyanpei@gmail.com}
\affiliation{%
  \institution{ARC Lab, Tencent PCG}
  \city{Beijing}
  \country{China}
}



\begin{abstract}

Synthesizing photorealistic 4D human head avatars from videos is essential for VR/AR, telepresence, and video game applications. Although existing Neural Radiance Fields (NeRF)-based methods achieve high-fidelity results, the computational expense limits their use in real-time applications. To overcome this limitation, we introduce \emph{BakedAvatar}, a novel representation for real-time neural head avatar synthesis, deployable in a standard polygon rasterization pipeline. Our approach extracts deformable multi-layer meshes from learned isosurfaces of the head and computes expression-, pose-, and view-dependent appearances that can be baked into static textures for efficient rasterization. We thus propose a three-stage pipeline for neural head avatar synthesis, which includes learning continuous deformation, manifold, and radiance fields, extracting layered meshes and textures, and fine-tuning texture details with differential rasterization. Experimental results demonstrate that our representation generates synthesis results of comparable quality to other state-of-the-art methods while significantly reducing the inference time required. We further showcase various head avatar synthesis results from monocular videos, including view synthesis, face reenactment, expression editing, and pose editing, all at interactive frame rates on commodity devices.
Source codes and demos are available on our project page: \textcolor{magenta}{\href{https://buaavrcg.github.io/BakedAvatar}{https://buaavrcg.github.io/BakedAvatar}}.

\end{abstract}



\begin{teaserfigure}
  \centering
  \includegraphics[width=\textwidth]{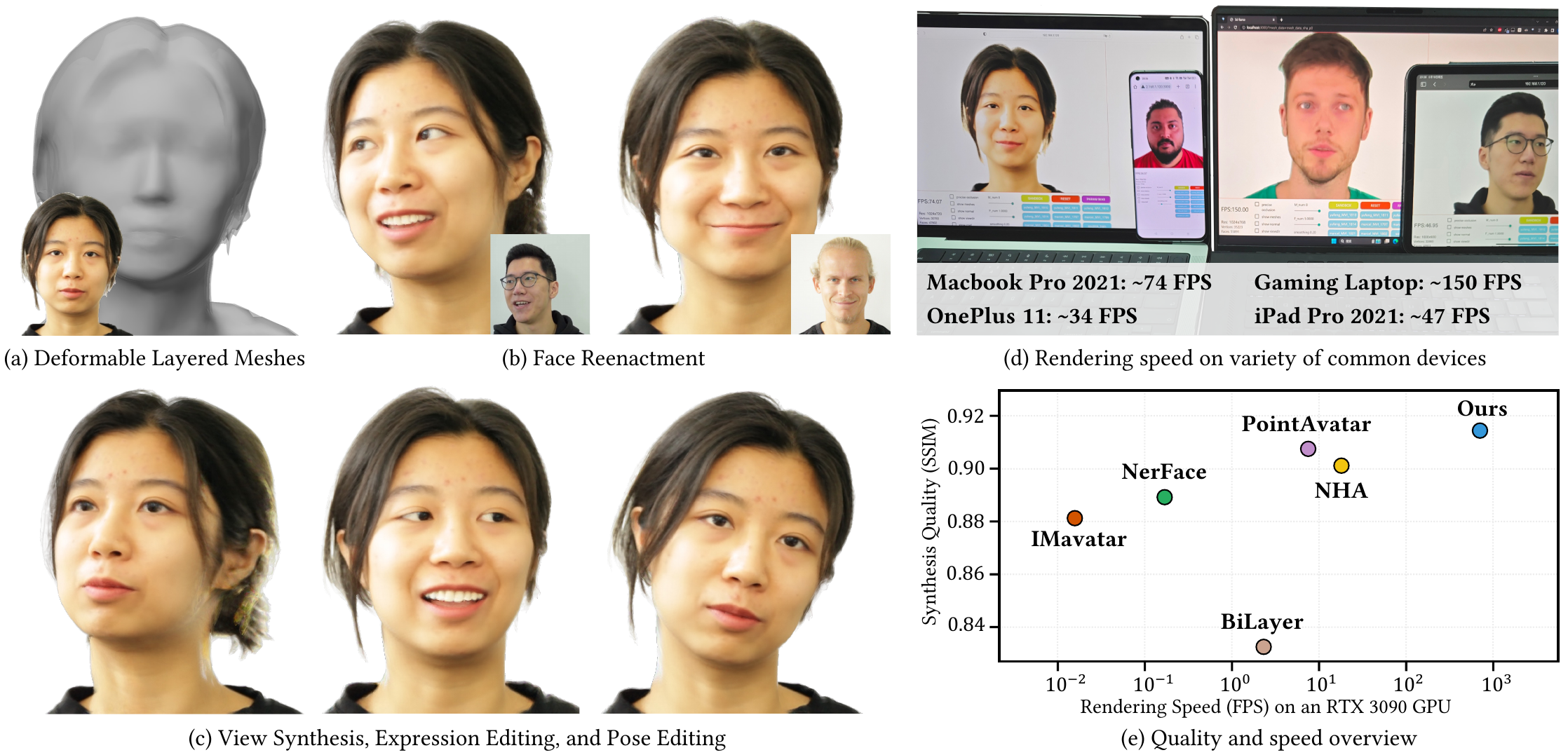}
  \caption{In \emph{BakedAvatar}, the learned neural fields of a head avatar are baked into (a) deformable layered meshes (and corresponding textures). This enables (b) face reenactment, (c) view synthesis, expression editing, and pose editing applications. It supports (d) real-time rendering on a variety of common devices with a standard polygon rasterization pipeline, and (e) outperforms alternatives in rendering speed while maintaining comparable or superior rendering quality.  }
  \label{fig:teaser}
\end{teaserfigure}


\maketitle

\section{Introduction}

The digitization of photorealistic 4D (or dynamic 3D) human heads has been a significant area of research in the fields of computer graphics and vision due to its importance in various applications such as VR/AR, telepresence, and video games. 
Given the complexity of human head shape, appearance, and movement, the precise replication of head avatar details from monocular videos presents a considerable challenge. 
Recently, the advent of neural rendering methods, particularly Neural Radiance Fields (NeRF)~\cite{mildenhall2020nerf}, has substantially improved the quality of novel view synthesis.
Subsequent applications of NeRF have accomplished 4D head avatar reconstruction by employing morphable facial models~\cite{3dmm, cao2013facewarehouse, li2017learning} with either direct expression and pose conditioning or learned linear blendable bases~\cite{gafni2021dynamic, gao2022reconstructing}. Other techniques involve the deformation of motion grids, meshes, or implicit functions~\cite{xu2023avatarmav, zielonka2023cvpr, zheng2022avatar}; or the learning of warpings to a canonical space~\cite{park2021nerfies, park2021hypernerf}.

Despite yielding photorealistic head avatar reconstructions, NeRF-based methods come with extensive computational demands. This is largely due to the dense ray-marching and implicit scene representation, which requires large neural networks to evaluate hundreds of samples along each ray for every pixel rendered. Consequently, their real-time application on commodity devices is significantly limited.
Recent advancements have sought to address this issue by converting a trained NeRF into explicit or hybrid representation for fast inference, such as multi-plane images (MPIs)~\cite{wizadwongsa2021nex}, sparse voxel grids~\cite{hedman2021baking}, meshes~\cite{Chen_2023_CVPR, yariv2023bakedsdf}, and tri-planes~\cite{reiser2023merf}. The conversion process, often called ``baking'', essentially pre-computes implicit representation and stores them in data structures that can be queried with minimal computation. However, these methods primarily focus on speeding up the inference of static scenes and do not readily extend to the more complex challenge of reconstructing dynamic objects.

The primary goal of this paper is to develop a representation that is optimized for 4D head avatar rendering, capable of running in real-time even on commodity devices such as mobiles. This introduces several constraints, including efficient implementation across devices, moderate storage demands, and the capability to animate based on user input expression or pose conditions. One might consider a mesh-based representation, the best supported primitive in the modern graphics pipeline. However, conventional mesh-based solutions primarily record appearance and compute shading at object surfaces, which significantly differs from volume-based NeRF. Although there are methods~\cite{oechsle2021unisurf, wang2021neus, yariv2021volume} that resort to surface-based reconstruction, we find them unsatisfactory for human head avatars, as pure surfaces struggle to capture semi-transparent and fine-scale structures such as hair. Furthermore, recovering detailed surface geometry would necessitate a highly intricate mesh, which could undermine the rendering performance.

Building upon the observations discussed previously, we present \emph{BakedAvatar}, a novel representation to achieve person-specific photorealistic head avatar synthesis harnessing rasterization, aiming to match the volume rendering quality provided by NeRF.
While NeRF inherently demands a large number of sample points along the ray to produce impeccable results, recent advancements~\cite{neff2021donerf, Attal_2023_CVPR, deng2022gram, chen2021mvsnerf, lin2022efficient} have demonstrated the feasibility of a substantial reduction in sampling points, as long as the locations of key sampling points are available. For instance, GRAM~\cite{deng2022gram} employs a learned manifold that guides the radiance sampling, defining implicit isosurfaces and utilizing only their intersections with camera rays for ray integration.
The crux of our proposed method is to similarly learn a manifold, but one that closely envelops the human head surface, allowing us to significantly minimize the number of sampling points. We then proceed to extract multiple mesh proxies from this learned manifold, enabling direct calculation of ray-manifold intersections. Our proposed representation shares similarities with MPIs~\cite{single_view_mpi}, but is warped around the human head. This unique configuration accommodates larger deformations than conventional MPIs can handle.

We then animate the proposed layered mesh representation using a morphable model. Specifically, our approach adopts FLAME~\cite{li2017learning}, a parametric head model that offers control over expression and pose. By learning a FLAME deformation field, we are able to export the derived blendshapes and skinning weights as vertex attributes. Furthermore, we propose to bake expression-, pose- and view-dependent appearance as a combination of linearly blendable texture bases. In addition, we develop a lightweight appearance decoder that runs efficiently during the shading process. Once trained from monocular videos, our representation enables real-time synthesis of 4D head avatars on consumer-grade devices.

To summarize, our major contributions include:
\begin{itemize}[leftmargin=18pt,topsep=3pt,itemsep=2pt]
  \item We present \emph{BakedAvatar}, a novel representation for 4D head avatars that bakes neural fields into deformable layered meshes and corresponding textures, designed for a rasterization pipeline.
  \item We propose a multi-layer mesh proxy that approximates the volumetric rendering of human heads. These meshes are rigged with FLAME-based morphing whose weights are extracted from a learned deformation field.
  \item We bake expression-, pose- and view-dependent appearance into linear blendable texture bases, and compose them with a light-weighted appearance decoder running in pixel shaders.
  \item Experimental results show that our method synthesizes head avatars of comparable quality to state-of-the-art (SOTA) methods while being significantly faster. We demonstrate that our representation achieves real-time rendering on commodity devices including laptops, tablets, and mobile phones, while enabling interactive expression and pose editing, as well as face reenactment.
\end{itemize}

\begin{figure*}[t!]
  \centering
  \includegraphics[width=1.0\linewidth]{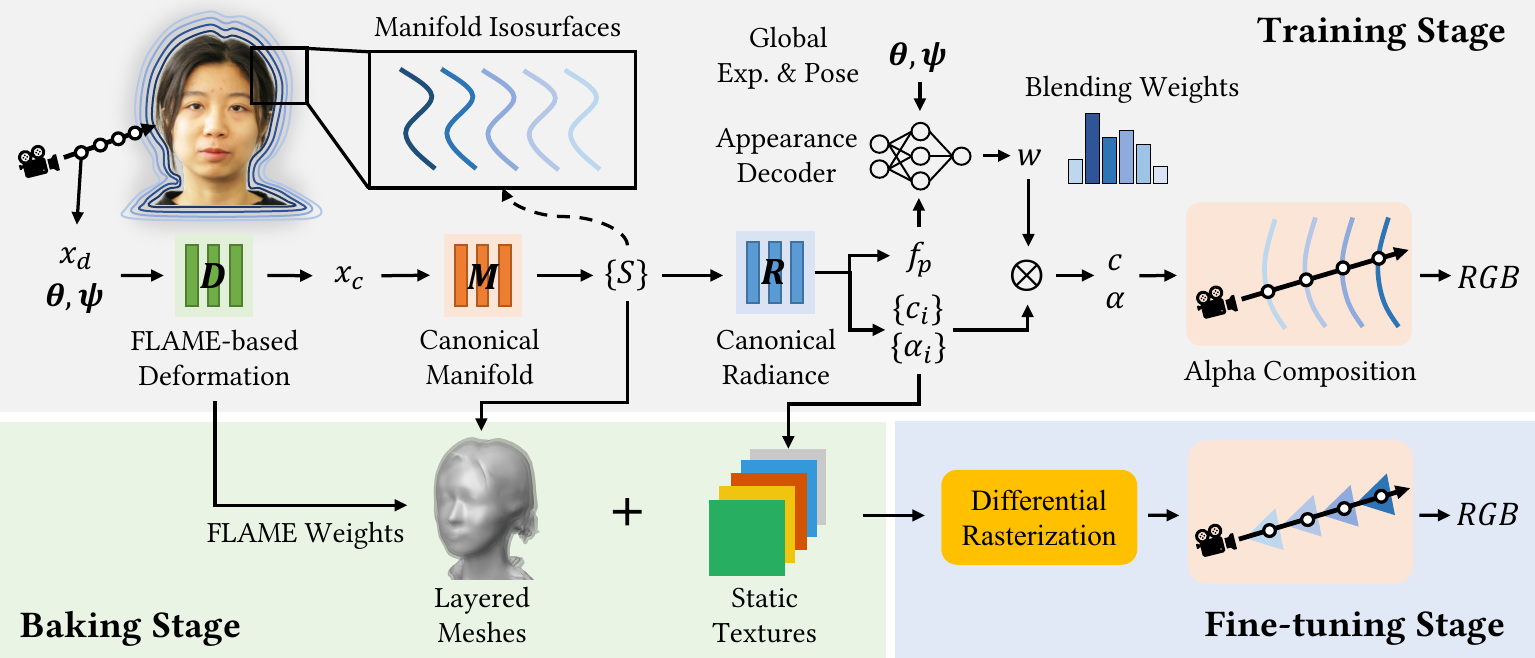}
  \caption{Overview of our three-stage pipeline. In the first stage, we learn three implicit fields in canonical space for FLAME-based deformation, isosurface geometry, as well as multiple radiance bases and a position feature that are combined by the light-weighted appearance decoder. In the second stage, we bake these neural fields as deformable multi-layer meshes and multiple static textures. In the third stage, we use differential rasterization to fine-tune the baked textures.}
  \label{fig:pip}
\end{figure*}

\section{Related Work}

\subsection{4D Head Avatar Synthesis}

Extensive research has been conducted on synthesizing 4D human head avatars. Image-based approaches reconstruct human heads from portraits or monocular videos without any 3D representation. These methods include encoder-decoder models~\cite{kim2018deep, thies2019deferred, zakharov2020fast}, warping fields~\cite{Siarohin_2019_NeurIPS}, and generative adversarial networks with explicit control~\cite{wang2023styleavatar}. Although Deep Video Portraits~\cite{kim2018deep, wen2020photorealistic} can run on commodity devices almost in real-time, image-based methods still face challenges with view inconsistencies.  Mesh-based methods output 3D human head meshes based on portraits or videos along with geometry prior~\cite{li2017learning}, mitigating the above weaknesses. Some reconstruct a morphable head model without textures from a portrait~\cite{Feng:SIGGRAPH:2021, MICA:ECCV2022}, while others utilize a hybrid representation to model the morphable head model and the texture space~\cite{grassal2022neural, Khakhulin2022ROME, Bai_2023_CVPR}. Although morphable models are convenient to animate and stable in the classical rendering pipeline, they require a trade-off between efficiency and quality. Recently, NeRF-based methods have gained attention. Deformable NeRFs have been explored for dynamic scenes~\cite{pumarola2021d, park2021nerfies, park2021hypernerf,liu2022devrf}, which can be applied to human heads. 
To achieve accurate animations, NeRFs are usually combined with 3D Morphable Models~\cite{gafni2021dynamic, gao2022reconstructing, zheng2022avatar, xu2023avatarmav} or a generative framework~\cite{wu2022anifacegan, xu2023latentavatar, yu2023nofa}. However, these methods require dense ray sampling and evaluation of large neural networks to predict attributes such as vertex displacement, radiance, density, surface position, etc. The significant computational requirements make real-time rendering prohibitive on consumer-grade devices, especially for mobiles. More recent works are gradually approaching interactive rendering on desktop GPUs, with methods such as mesh-guided deformation~\cite{zielonka2023cvpr} and real-time inversion of a trained 3D portrait generator~\cite{trevithick2023real}. Our approach takes a further step that directly renders multi-layer meshes, achieving real-time framerates even on mobile devices.

\subsection{Hybrid NeRF Representation for Accelerated Inference}

Minimizing the computational expense of NeRFs for rapid inference using explicit or hybrid scene representations is an active research direction. One solution is to split large Multi-Layer Perceptrons (MLPs) in NeRF into small chunks and cache them independently~\cite{garbin2021fastnerf} or store them in smaller MLPs~\cite{reiser2021kilonerf}. Sparse voxel structures such as octrees can efficiently represent a scene by skipping empty spaces~\cite{liu2020neural}, encoding appearance using baked spherical harmonics~\cite{yu2021plenoctrees, fridovich2022plenoxels}, or storing features which can be decoded with a tiny deferred shading network to produce specular appearance~\cite{hedman2021baking}. Image-based representations like MPIs~\cite{wizadwongsa2021nex} or triplane decomposed volumes~\cite{reiser2023merf}, can store a scene with less memory. Mesh-based representation is rarely used for NeRFs as shading only occurs on the surface. Still, recent work has converted NeRFs to meshes for rasterization pipeline with an optimized polygon soup~\cite{Chen_2023_CVPR} or signed distance fields~\cite{yariv2023bakedsdf}. Neumesh~\cite{yang2022neumesh} also bakes NeRF to a triangle mesh for editing. The exploration of hybrid mesh-volume representation~\cite{guo2023vmesh} is also underway. Dynamic scenes can be parameterized using Fourier-PlenOctrees~\cite{wang2022fourier} or 4D voxel deformation fields~\cite{liu2022devrf}; however, these approaches necessitate substantial memory consumption and are unsuitable for animatable head avatars.
Our work focuses on real-time inference of an animatable head avatar by baking
neural fields into deformable layered meshes and corresponding textures without using large MLPs or ray marching-based radiance samplings.

\section{Method}

Our main idea is to learn multi-layer meshes that approximate volume rendering in NeRFs. These layered meshes should possess the ability to deform in response to head pose and expression variations while also accurately capturing dynamic changes in appearance caused by micro-expressions. Further, the neural fields associated with a head avatar can be baked into the deformable layered meshes and corresponding textures for real-time rendering.
To this end, we develop a three-stage method to train the person-specific head avatar. First, we reconstruct the 4D head avatar with implicit fields designed for baking (Sec.~\ref{stage1}). Second, we extract the deformable layered meshes and bake the radiance field into static textures for real-time rendering (Sec.~\ref{stage2}). Finally, we fine-tune this representation using differential rasterization (Sec.~\ref{stage3}). The pipeline of our method is illustrated in Fig.~\ref{fig:pip}.

\subsection{Learning Deformable Manifold and Expression-, Pose-, View-Dependent Appearance} \label{stage1}

To acquire multi-layer meshes and enable real-time 4D head avatar synthesis from monocular videos, we first reconstruct the head avatar with an implicit representation. Specifically, we learn three continuous fields: a manifold field for multi-layer geometry, a forward deformer for FLAME-based deformation, and a radiance field for dynamic appearance conditioned on expression, pose, and view.

\subsubsection{Canonical Manifold Field} \label{canonical_manifold}

We learn proxy layered surfaces to constrain radiance sampling points, with the surfaces represented as predefined level sets of a manifold field. The manifold field is established in canonical space, which is shared by all expressions and poses to streamline the training process. In accordance with GRAM~\cite{deng2022gram}, we employ an MLP to implement the manifold field, taking a canonical position $\mathbf{x}_c$ as input and generating a scalar level $s$, expressed as 
\begin{equation}
    \mathcal{M}: \mathbf{x}_c \in \mathbb{R}^3 \to s \in \mathbb{R}.
\end{equation}

Radiance accumulation occurs on the learned layered surfaces, which are given by canonical space intersections of the ray and implicit isosurfaces $\mathcal{S}=\{S_i\}$ of predefined level sets $\Gamma=\{l_i\}$. Let $\mathbf{\hat{x}}_c$ be a canonical intersection point, the $i$-th layer surface is denoted as $S_i = \{ \mathbf{\hat{x}}_c | \mathcal{M}(\mathbf{\hat{x}}_c) = l_i \}$.
Undesirable intersections among various layers are substantially avoided, as they are defined as isosurfaces of the manifold level sets, which assist in extracting well-structured meshes as described in Sec.~\ref{stage2}. We implement the canonical manifold using a coordinate-based MLP and adopt geometry initialization~\cite{atzmon2020sal} to prevent training divergence. Furthermore, we employ a low position encoding frequency to promote smooth geometry, facilitating the extraction of simpler meshes.

\subsubsection{FLAME-Driven Forward Deformer} \label{deformer}

To achieve real-time manipulation of the layered meshes based on head expression and pose changes, we adopt the FLAME deformation model~\cite{li2017learning}, which allows us to compute deformations with per-vertex blendshape and linear blend skinning (LBS) operations on the extracted meshes. Unlike backward deformation used in dynamic NeRFs~\cite{pumarola2021d, park2021nerfies, park2021hypernerf}, the forward deformer transforms a canonical isosurface to a deformed one that intersects with camera rays. 
Following IMavatar~\cite{zheng2022avatar}, the blendshape and LBS weights of a point $\mathbf{x}_c$ in canonical space are predicted with a deformation field $\mathcal{D}: \mathbf{x}_c \in \mathbb{R}^3 \to \mathcal{E},\mathcal{P},\mathcal{W}$,
where $\mathcal{E} \in \mathbb{R}^{n_e \times 3}$ are $n_e=50$ expression blendshapes, $\mathcal{P} \in \mathbb{R}^{n_p \times 9 \times 3}$ are $n_p=4$ pose correctives, and $\mathcal{W} \in \mathbb{R}^{n_j}$ are $n_j=5$ LBS bone weights, all at canonical points. The location of each deformed point $\mathbf{x}_d$ from canonical point $\mathbf{x}_c$ is computed using the FLAME deformation formula:
\begin{equation}
    \mathbf{x}_d = LBS(\mathbf{x}_c + B_E(\psi; \mathcal{E}) + B_P(\theta; \mathcal{P}), J(\mathbf{x}_c + B_E(\psi; \mathcal{E})),\theta,\mathcal{W}),
\end{equation}
where $\psi$ and $\theta$ represent the FLAME expression and pose coefficients for the current frame, while $B_E(\cdot)$ and $B_P(\cdot)$ compute the expression and pose offsets using the corrective blendshape bases $\mathcal{P}$ and $\mathcal{E}$. $J(\cdot)$ serves as the joint regressor.

As we learn a deformation field in canonical space, we follow the iterative root finding scheme in SNARF~\cite{chen2021snarf} to retrieve the canonical correspondence of ray sampling points in the deformed space. More details can be found in the appendix.

\begin{figure}[!t]
  \centering
  \includegraphics[width=\linewidth]{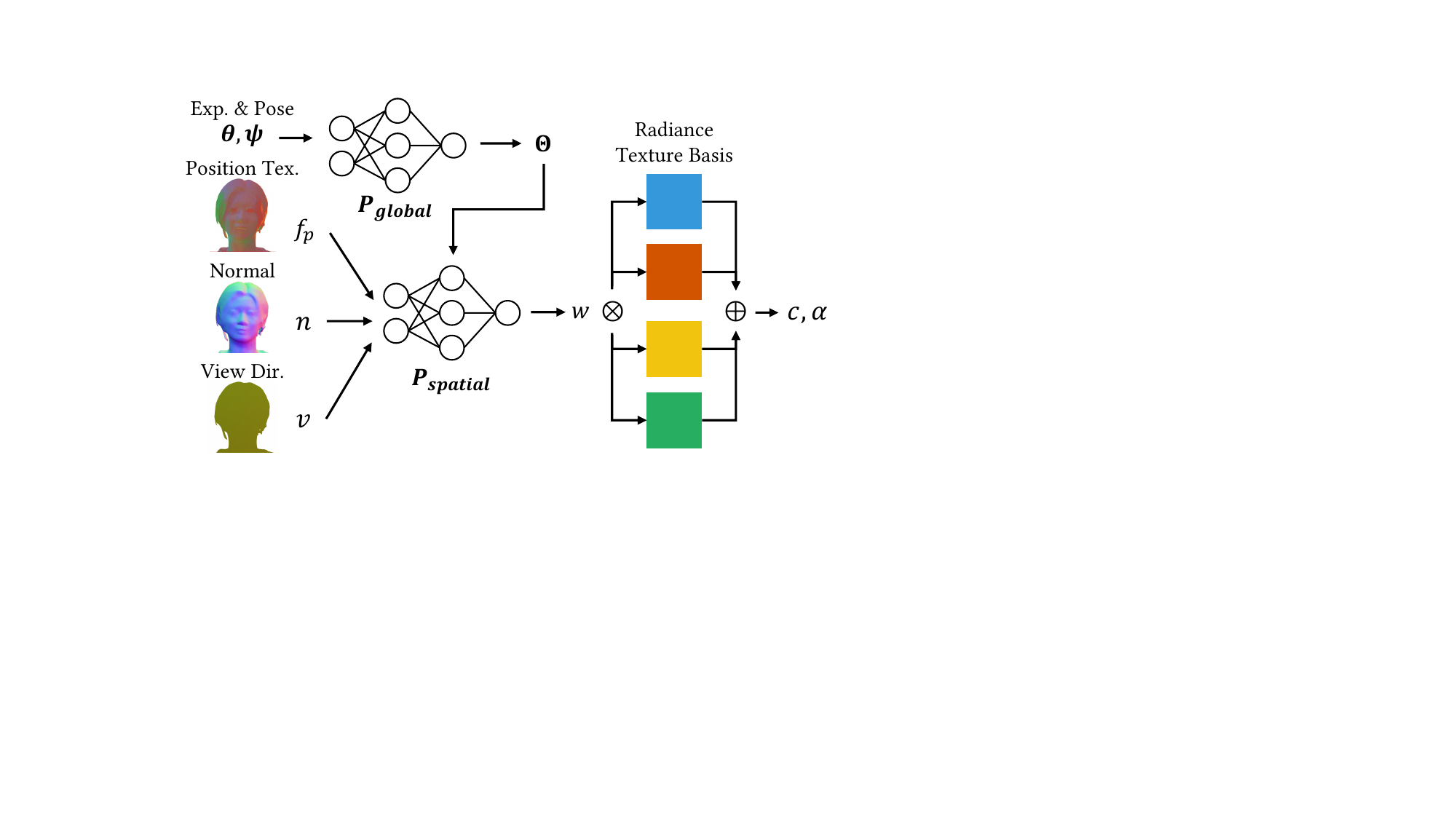}
  \caption{The light-weighted appearance decoder $P_\text{app}$ consists of a global MLP and a spatial MLP. The global MLP takes inputs of global expression and pose conditions and produces the weights of spatial MLP. The spatial MLP takes inputs of position texture, normal, and view direction, which vary spatially in screen space, and produces normalized blending weights. These weights are then combined with multiple radiance texture bases to generate color and alpha values of the current mesh layer.}
  \label{fig:decoder}
\end{figure}

\subsubsection{Expression-, Pose-, and View-Dependent Radiance Field}

The appearance of a head avatar is highly dynamic and complex, influenced by factors such as viewing direction, pose and expression, etc. To model the expression-, pose, and view-dependent effects, the radiance field takes input from FLAME expression coefficients $\psi \in \mathbb{R}^{n_e}$ and rotation of jaw and eye joints $\theta \in \mathbb{R}^{3 \times 3}$. Additionally, we condition the radiance field with the spatially varying normal $\mathbf{n} \in \mathbb{R}^3$ and view direction $\mathbf{v} \in \mathbb{R}^3$, both in the camera space. The radiance field predicts RGB color $\mathbf{c} \in \mathbb{R}^3$ and occupancy $\alpha \in [0,1]$ in canonical space, which is denoted as 
\begin{equation}
    \mathcal{R}: (\mathbf{x}_c,\mathbf{v},\mathbf{n},\mathbf{\theta},\mathbf{\psi}) \to (\mathbf{c},\alpha) \in \mathbb{R}^4.
\end{equation}

To ensure real-time inference performance and avoid evaluating large MLPs, we represent this radiance field using multiple linear blendable bases, where each basis can be easily baked into a static texture for efficient querying. Concretely, assuming there are $n_T$ texture bases, the radiance field then predicts bases' colors $\{\mathbf{c}_1, \mathbf{c}_2, \cdots, \mathbf{c}_{n_T}\}$ and occupancies $\{\mathbf{\alpha}_1, \mathbf{\alpha}_2, \cdots, \mathbf{\alpha}_{n_T}\}$, as well as a shared $d_p$-dimensional position feature $\mathbf{f}_p \in \mathbb{R}^{d_p}$ in canonical space, denoted as 
\begin{equation}
    \mathcal{R}_\text{basis} : \mathbf{x}_c \in \mathbb{R}^3 \to (\{\mathbf{c}^i\}, \{\mathbf{\alpha}^i\}, \mathbf{f}_p) \in \mathbb{R}^{4 \times n_T + d_p}.
\end{equation}

The position feature contains high-frequency textural information which can be recorded in a static texture. We use an appearance decoder $P_\text{app}$ to project the global and spatial varying conditions $(\mathbf{f}_p,\mathbf{v},\mathbf{n},\mathbf{\theta},\mathbf{\psi})$ into softmax normalized weights $\mathbf{w} \in \mathbb{R}^{n_T}$ for blending multiple radiance bases.
As shown in Fig.~\ref{fig:decoder}, the decoder is implemented as a tiny hyper-network that consists of two parts: $P_\text{global}$, mapping global condition to the weights of $P_\text{spatial}$, and $P_\text{spatial}$, predicting blending weights from spatial varying conditions:
\begin{equation}
\begin{cases}
    P_\text{global} : (\mathbf{\theta},\mathbf{\psi}) \to \Theta_\text{spatial} \\
    P_\text{spatial} : (\mathbf{f}_p,\mathbf{v},\mathbf{n}; \Theta_\text{spatial}) \to \mathbf{w} \in \mathbb{R}^{n_T}.
\end{cases}
\end{equation}

The hyper-network efficiently reduces the computation related to global conditions, as the global MLP requires evaluation only once per frame, while the spatial MLP is evaluated at each ray-manifold intersection. The final radiance is derived as a linear combination of radiance from all bases: 
\begin{equation}
    (\mathbf{c},\mathbf{\alpha}) = \sum_{i\in n_T} \mathbf{w}_i (\mathbf{c_i},\mathbf{\alpha}_i).
\end{equation}

\subsubsection{Training Strategy}
We employ non-rigid ray marching to identify the canonical manifold isosurfaces and collect the radiance and occupancy at their intersections along the ray using alpha compositing:
\begin{equation}
C(r)=\sum_{j=1}^M \prod_{k<j}(1-\alpha(\mathbf{x}_j))\alpha(\mathbf{x}_j)\mathbf{c}(\mathbf{x}_j)
\end{equation}
where $\alpha(\mathbf{x}_j)$ is the opacity at position $\mathbf{x}_j$, and $\mathbf{c}(\mathbf{x}_j)$ is the color at position at position $\mathbf{x}_j$. Note that this formulation is different from the original rendering formula in NeRF, as we are using multiple surfaces instead of sampling points in a volume.
During training, we randomly sample a number of rays, and the geometry, deformation, and appearance fields are jointly optimized in an end-to-end manner. This process is supervised by image-based losses and pseudo deformation ground truth extracted from the off-the-shelf expression capture module DECA~\cite{Feng:SIGGRAPH:2021}. We initially apply a photometric loss on the rendered color of each ray $r$ where a ray-manifold intersection is found:
\begin{equation}
    \mathcal{L}_\text{rgb} = \frac{1}{\vert R \vert} \sum_{r \in R} \Vert \mathbf{C}_r - \mathbf{C}_r^\text{GT} \Vert_{1},
\end{equation}
where $R$ is the set of rays intersected with the isosurfaces $\mathcal{S}$, $\mathbf{C}_r$ is the rendered color, $\mathbf{C}_r^\text{GT}$ is the corresponding ground truth color.

We also apply a weighted mask loss on all manifold layers, encouraging outer layers to match the ground truth mask with higher weights and gradually reducing the weight towards inner layers:
\begin{equation}
    \mathcal{L}_\text{mask} = \frac{1}{n_T} \sum_{i}^{n_T} \frac{exp(-\gamma \Vert l_i \Vert)}{\vert \overline{R}_i \vert} \sum_{r \in \overline{R}_i} CE(-s^i_r, O_r)
\end{equation}
where $\overline{R}$ is the set of non-surface rays, $O_r$ is the ground truth mask, $s^i_r$ is the predicted manifold scalar at ray intersection with the $i$-th isosurface, $CE(\cdot)$ is the binary cross entropy loss, $l_i$ is the predefined level, and $\gamma=10$ is the attenuation factor. The ground truth mask is obtained with an off-the-shelf foreground estimator~\cite{ke2022modnet}.

Following~\cite{zheng2022avatar}, we supervise the deformation network with pseudo FLAME ground truth, and encourage a lower Manhattan distance between isosurfaces and FLAME surface for the skin region $R^\text{skin}$:
\begin{equation}
\begin{aligned}
    \mathcal{L}_\text{flame} &= \frac{1}{n_T} \sum_{i}^{n_T} \Big [ \sum_{r \in R_i} (\lambda_e \Vert \mathcal{E}_r - \mathcal{E}_r^\text{GT} \Vert_2 + \lambda_r \Vert \mathcal{P}_r - \mathcal{P}_r^\text{GT} \Vert_2 \\ &+ \lambda_w \Vert \mathcal{W}_r - \mathcal{W}_r^\text{GT} \Vert_2) + \sum_{r \in R^\text{skin}_i} \lambda_d \Vert \mathbf{x}_r - \mathbf{x}_r^\text{GT} \Vert_2 \Big ],
\end{aligned}
\end{equation}
where $\mathcal{E}_r$, $\mathcal{P}_r$, and $\mathcal{W}_r$ represent the predicted values of the deformation network, $\mathbf{x}_r$ is the position of the ray-isosurface intersection point, while $\mathcal{E}^\text{GT}_r$, $\mathcal{P}^\text{GT}_r$, $\mathcal{W}^\text{GT}_r$ and $\mathbf{x}^\text{GT}_r$ denote the pseudo ground truth weights and positions defined by the nearest FLAME vertices, $\lambda_e=1000$, $\lambda_r=1000$, $\lambda_w=0.1$, $\lambda_d=1$ are coefficients.

In summary, the overall loss function is:
\begin{equation}
    \mathcal{L}_\text{train} = \lambda_\text{rgb} \mathcal{L}_\text{rgb} + \lambda_\text{mask} \mathcal{L}_\text{mask} + \lambda_\text{flame} \mathcal{L}_\text{flame},
\end{equation}
where $\lambda_\text{rgb}=1$, $\lambda_\text{mask}=10$, $\lambda_\text{flame}=1$ are loss weights.

\begin{figure}[!t]
  \centering
  \includegraphics[width=\linewidth]{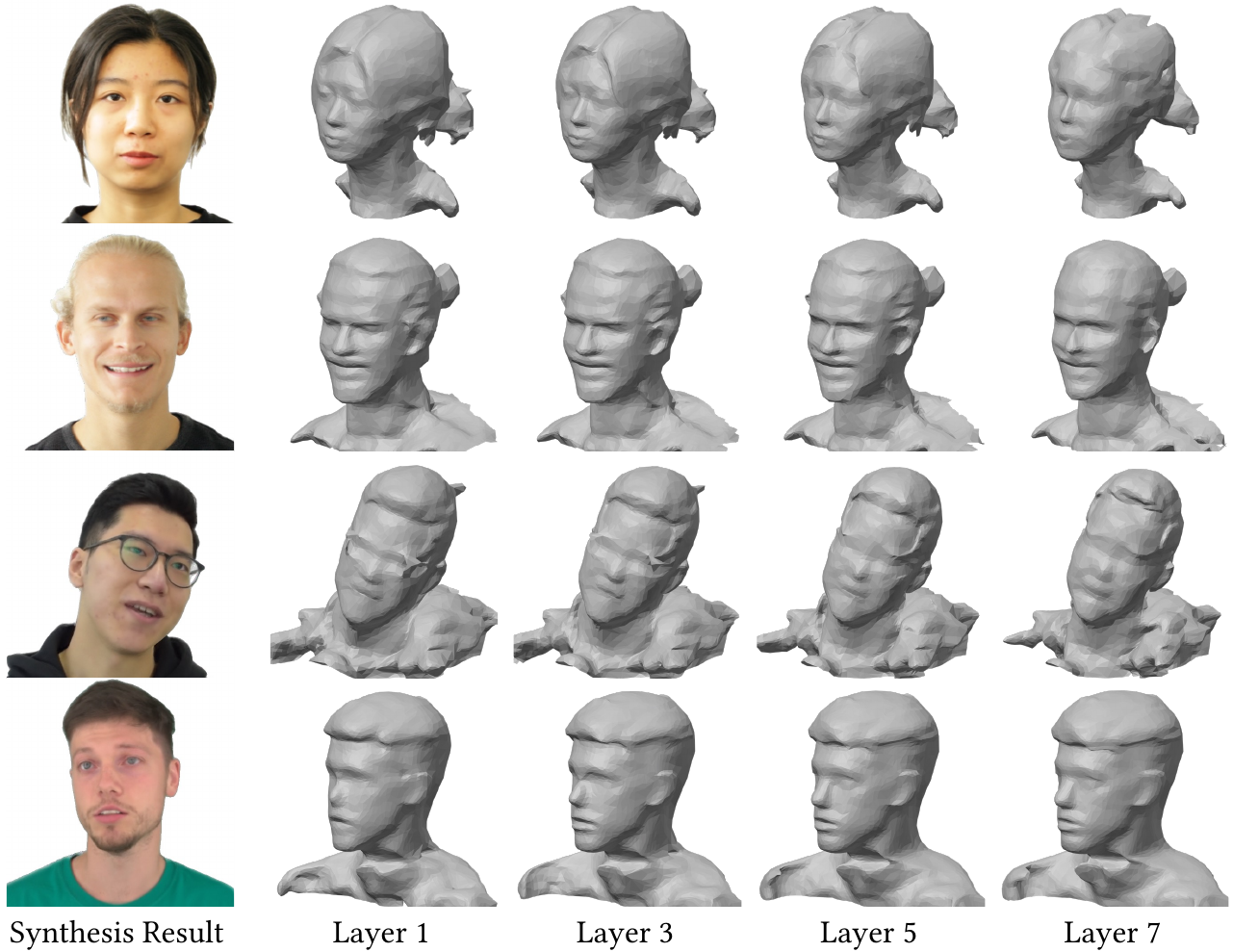}
  \caption{
  Visualization of extracted meshes under the given pose and expression, shown from outer layers (on the left) to inner layers (on the right). It can be seen that these multiple layers of meshes generally conform to the shape of the head. They also offer the essential intersection points needed to approximate the effects of volumetric rendering.
  }
  \label{fig:meshvis}
\end{figure}

\subsection{Computing Deformable Layered Meshes and Textures} \label{stage2}
Following the first stage, we bake the continuous implicit fields to layered meshes and multiple textures for real-time rendering. We begin with mesh extraction, using Marching Cubes~\cite{lorensen1987marching} to extract isosurfaces in canonical space for each predefined level. Since there are excess triangle faces in the smooth area, and the back region of the head is not visible during rendering, we apply a selective mesh simplification scheme using Quadric Edge Collapse Decimation~\cite{garland1997surface} to improve real-time inference performance. Specifically, we first reduce the face count to a target number and then further simplify faces with normal $\mathbf{n}$ that satisfy $\arccos(\mathbf{n} \cdot \mathbf{n}_\text{view}^T) > \theta_{max}$ to a target percentage, where $\mathbf{n}_\text{view}$ is set to the Z+ axis (the direction facing the head). Fig.~\ref{fig:meshvis} visualizes the representative deformable mesh layers. More details are provided in Sec.~\ref{mesh_precision} and Sec.~\ref{perfermance_measurement}. 

We bake deformation parameters onto extracted mesh vertex attributes to perform mesh morphing at inference time. For each mesh vertex in canonical space, we evaluate the deformation field to obtain the corresponding FLAME weights, which we store as vertex attributes. We also evaluate the normalized gradients of the manifold field as vertex normals.
We apply UV unwrapping using xaltas~\cite{xaltas} to acquire texture coordinates for each vertex and densely sample all texels for baking radiance bases and position features from $R_\text{basis}$. Specifically, for each texel, we uniformly sample 16 points and find their corresponding positions in canonical space. We then evaluate the radiance field to obtain a smooth estimation of colors $\{\mathbf{c}_1, \mathbf{c}_2, \dots, \mathbf{c}_{n_T}\}$ and occupancies $\{\alpha_1, \alpha_2, \dots, \alpha_{n_T}\}$ for $n_T$ radiance bases, as well as the position feature $\mathbf{f}_p \in \mathbb{R}^{d_p}$. We store these as $n_T$ RGBA textures and a deep texture with $d_p$ channels, all quantized to the range [0, 255] and stored as 8-bit unsigned integers.

\begin{figure}[!t]
  \centering
  \includegraphics[width=\linewidth]{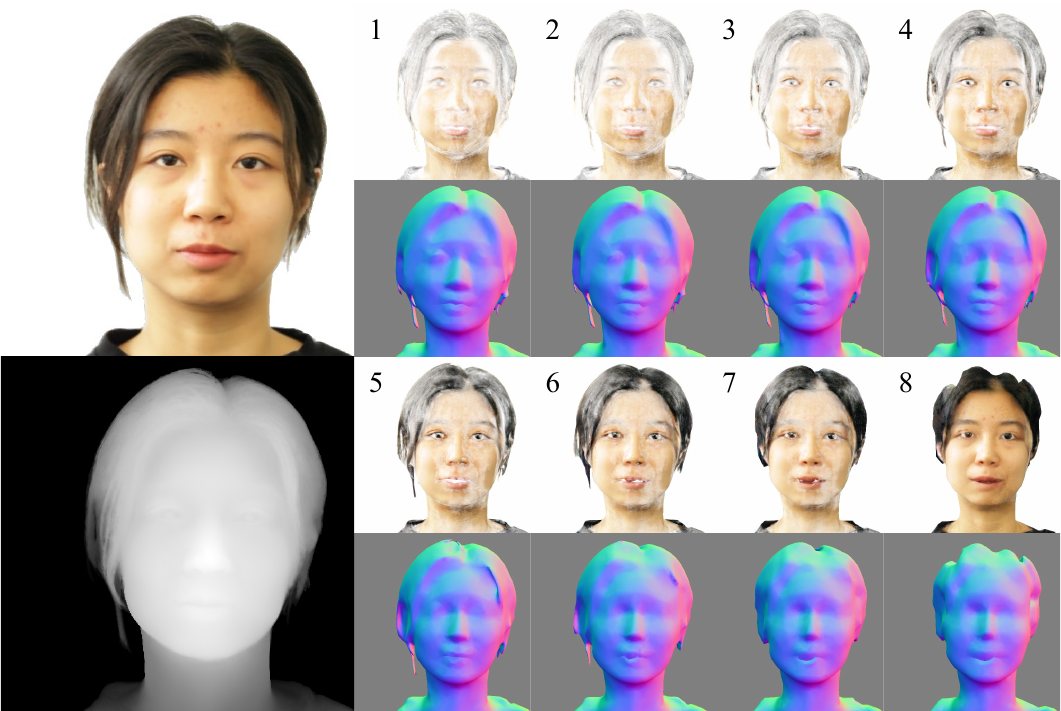}
  \caption{Visualization of the rendered color and depth, as well as the appearance and geometry of isosurfaces, from outer layers to inner layers.}
  \label{fig:layervis}
\end{figure}

\subsection{Fine-tuning Mesh Textures}
\label{stage3}
To minimize the errors introduced during the baking stage, we include a fine-tuning stage using differential rasterization to directly optimize mesh assets. As optimizing vertex positions of the discretized mesh is prone to local minima and may introduce undesirable intersections between manifold levels, we fix the mesh geometry and only optimize textures during fine-tuning. Fine-tuning textures helps eliminate artifacts at texture seams introduced in the UV parameterization process and refines appearance details that may not be well-reconstructed in the first stage. We render all mesh layers from inner levels to outer levels respectively, where the light-weighted MLP is evaluated at all rendered pixels to blend radiance texture bases, and finally combine them with standard alpha composition to produce the final color. We also optimize the camera translation and FLAME pose of each frame during the fine-tuning process, alleviating the errors of extracted camera extrinsic in datasets. We apply the same RGB loss from the first stage, and a VGG perceptual loss~\cite{johnson2016perceptual} since we now render the whole image instead of randomly sampled pixels:
\begin{equation}
    \mathcal{L}_\text{vgg} = \Vert F_\text{vgg}(\mathbf{C}) - F_\text{vgg}(\mathbf{C}^\text{GT}) \Vert,
\end{equation}
where $\mathbf{C}$ and $\mathbf{C}^\text{GT}$ are the rendered and ground truth images respectively, $F_\text{vgg}(\cdot)$ outputs features from the first four convolution layers of a pre-trained VGG network. The total loss for fine-tuning is:
\begin{equation}
    \mathcal{L}_\text{finetune} = \lambda_\text{rgb} \mathcal{L}_\text{rgb} + \lambda_\text{vgg} \mathcal{L}_\text{vgg},
\end{equation}
where $\lambda_\text{rgb}=1$, $\lambda_\text{vgg}=0.1$.

\subsection{Real-Time Rendering} \label{RTR}

We implement an interactive viewer with Javascript and WebGL2 on an HTML webpage. Our final representation for real-time rendering consists of four components: multi-layer meshes with vertex attributes (i.e., UVs, normals, FLAME weights), position and radiance texture atlases for each mesh, vertices and joint regressor of the FLAME template mesh, and weights of the appearance decoder. We compute blendshape offsets and LBS weights of each vertex from global expression and pose parameters in the vertex shader. The spatial MLP runs in the fragment shader, which calculates the blending coefficients of multiple radiance bases for all shaded pixels of each mesh layer in parallel. Specifically, the real-time rendering process includes three steps: (1) using expression and pose parameters at the current frame to calculate the global pose correctives and LBS bone transformations from the FLAME template mesh; (2) running the global MLP to produce dynamic weights of the spatial MLP, which are stored in a texture and then passed to the fragment shader; (3) drawing the layered meshes from inner to outer layers to ensure correct ordering in alpha composition (see Fig.~\ref{fig:layervis}). 
By harnessing the optimized rasterization pipeline of modern GPUs, our interactive viewer renders high-quality, dynamic head avatars in real time.

\begin{table}[!t]
  \caption{Quantitative Comparisons against SOTA methods.
  }
  \label{tab:comparison}
  \resizebox{\columnwidth}{!}{%
  \begin{tabular}{lP{2.2em}P{2.2em}P{2.2em}P{2.2em}P{1.6em}P{2.2em}}
    \toprule
    & L1$\downarrow$ & PSNR$\uparrow$ & SSIM$\uparrow$ & LPIPS$\downarrow$ & KD$\downarrow$ & FPS$\uparrow$\\
    \midrule
    BiLayer~\shortcite{zakharov2020fast} & $0.0765$ & $15.39$ & $0.830$ & $0.172$ & $5.77$ & $3.39$\\
    NerFace~\shortcite{gafni2021dynamic} & $0.0219$ & $25.68$ & $0.888$ & $0.112$ & $5.44$ & $0.15$\\
    IMavatar~\shortcite{zheng2022avatar} & $0.0242$ & $24.18$ & $0.881$ & $0.118$ & $5.13$ & $0.02$\\
    PointAvatar~\shortcite{Zheng2023pointavatar} & $0.0180$ & $27.42$ & $0.907$ & $0.060$ & $5.51$ & $8.53$\\
    Ours & $0.0147$ & $28.66$ & $0.917$ & $0.057$ & $4.97$ & $804$\\
    \midrule
    NHA~\shortcite{grassal2022neural} & $0.0204$ & $23.85$ & $0.902$ & $0.078$ & $6.08$ & $12.30$\\
    Ours (w/o cloth) & $0.0108$ & $30.93$ & $0.943$ & $0.045$ & $4.97$ & $831$\\
  \bottomrule
\end{tabular}}
\end{table}

\section{Implementation Details}

We select 8 manifold layers for manifold learning, with predefined levels sampled evenly from $-0.2$ to $0$, and use $n_T = 16$ and $d_p = 8$ for appearance learning.
In the baking stage, we set the target number of faces of each mesh layer to $10,000$, and further simplify backward-facing triangles with $\theta_{max} = 120^{\circ}$. Position feature and radiance bases are sampled and stored in $1024 \times 1024$ sized images. 
During fine-tuning, we utilize nvdiffrast~\cite{Laine2020diffrast} for differential rasterization on the layered meshes, with training frames scaled to $256\times256$ and $512\times512$ in the training and fine-tuning stages, respectively. Network architecture details are provided in the supplementary material.
We implement the training procedure in PyTorch~\cite{paszke2019pytorch}, using Adam~\cite{kingma2014adam} optimizer with an initial learning rate set to $3 \times 10^{-4}$, and decays by a factor of $0.5$ every 20k iterations. It takes about a day to train our model for a subject on a single RTX 3090 GPU.
We recognize that optimizing training time~\cite{muller2022instant, chen2023fast} is beyond the scope of this paper, as our primary objective is to achieve real-time inference and rendering.

\section{Experiments}

We conduct experiments using monocular videos of 8 subjects gathered from public datasets of PointAvatar~\cite{Zheng2023pointavatar}, NHA~\cite{grassal2022neural}, NerFace~\cite{gafni2021dynamic}, and two Youtube videos (\href{https://www.youtube.com/watch?v=ceZw5ZhIqwE}{ceZw5ZhIqwE} and \href{https://www.youtube.com/watch?v=-mIaZPqoOYw}{-mIaZPqoOYw}).
The training dataset contains 1,000-4,000 frames, while the test dataset includes frames with novel expressions and poses.
We assess our method by examining the image quality of self-reenactment and controlled head avatar synthesis, and the rendering speed on diverse devices, along with an ablation study.

\begin{figure*}[t!]
  \centering
  \includegraphics[width=0.98\textwidth]{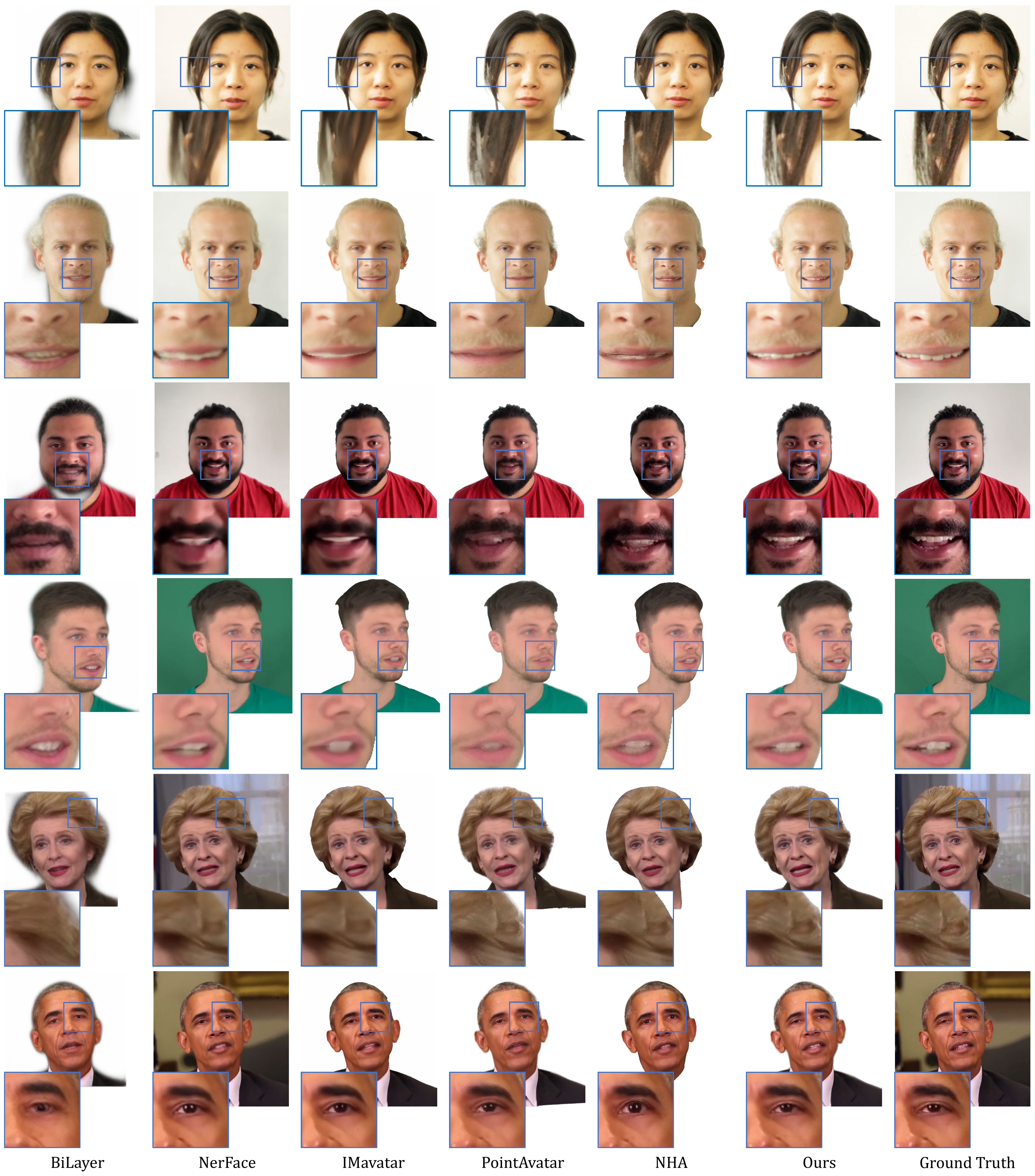}
  \vspace{-1.0mm}
  \caption{
  Comparisons of self-reenactment results. BiLayer~\cite{zakharov2020fast} fails to faithfully preserve the identity in the results. NerFace~\cite{gafni2021dynamic} and IMavatar~\cite{zheng2022avatar} generate blurry results due to the smooth nature of coordinate-based MLP. While NHA~\cite{grassal2022neural} generates sharp textures, it struggles to reconstruct semi-transparent thin structures such as hair. PointAvatar~\cite{Zheng2023pointavatar} generates volumetric effects well but fails to capture details of the mouth interior. 
  In contrast, our approach excels in capturing finer details, particularly in complex structures like hair and mouth.
  Ground truth videos \emph{Stabenow\textcopyright Senate Democrats} and \emph{Obama\textcopyright The Obama White House} in the last two rows are from public domain.
  }
  \vspace{-3.0mm}
  \label{fig:comparison}
\end{figure*}

\subsection{Comparison of Quality and Speed} \label{comparison}

\begin{table}[!t]
  \caption{Quantitative comparison of results after each stage.}
  \label{tab:comparasion_stage}
  \resizebox{0.96\columnwidth}{!}{%
  \begin{tabular}{lP{2.2em}P{2.2em}P{2.2em}P{2.2em}P{1.6em}P{2.2em}}
    \toprule
    & L1$\downarrow$ & PSNR$\uparrow$ & SSIM$\uparrow$ & LPIPS$\downarrow$ & KD$\downarrow$ \\
    \midrule
    Stage-1 (Training) & $0.0148$ & $28.14$ & $0.919$ & $0.085$ & $4.94$ \\
    Stage-2 (Baking) & $0.0170$ & $27.87$ & $0.915$ & $0.087$ & $5.02$ \\
    Stage-3 (Fine-tuning) & $0.0147$ & $28.66$ & $0.917$ & $0.057$ & $4.97$ \\
  \bottomrule
\end{tabular}}
\end{table}

\begin{figure*}[t!]
  \centering
  \includegraphics[width=\textwidth]{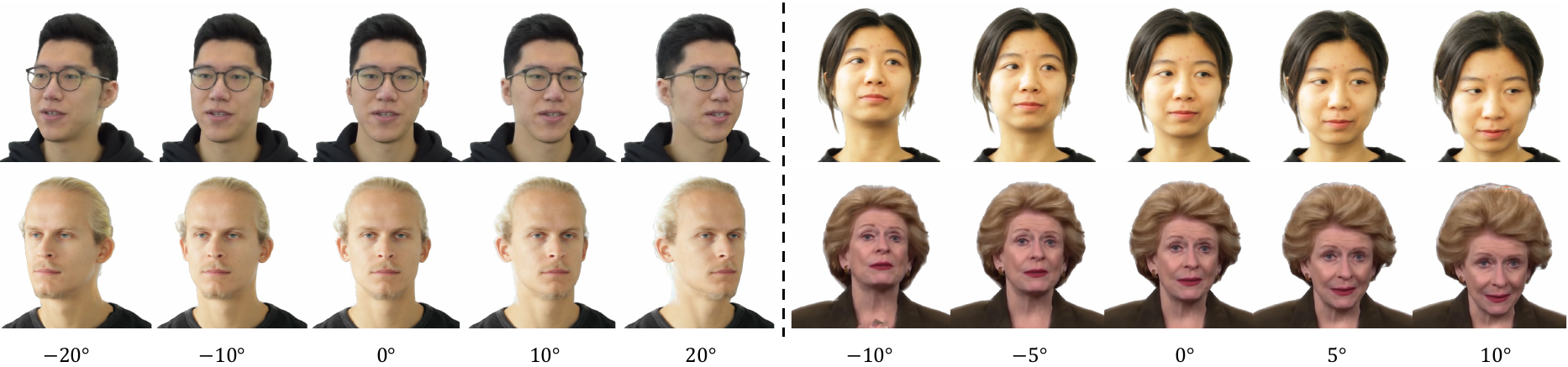}
  \vspace{-1.0mm}
  \caption{
  Novel view synthesis results by changing yaw (left) and pitch (right) angles. 
  The bottom right figure is from \emph{Stabenow\textcopyright Senate Democrats}.
  }
  \label{fig:view_synthesis}
\end{figure*}

\begin{figure*}[t!]
  \centering
  \includegraphics[width=\textwidth]{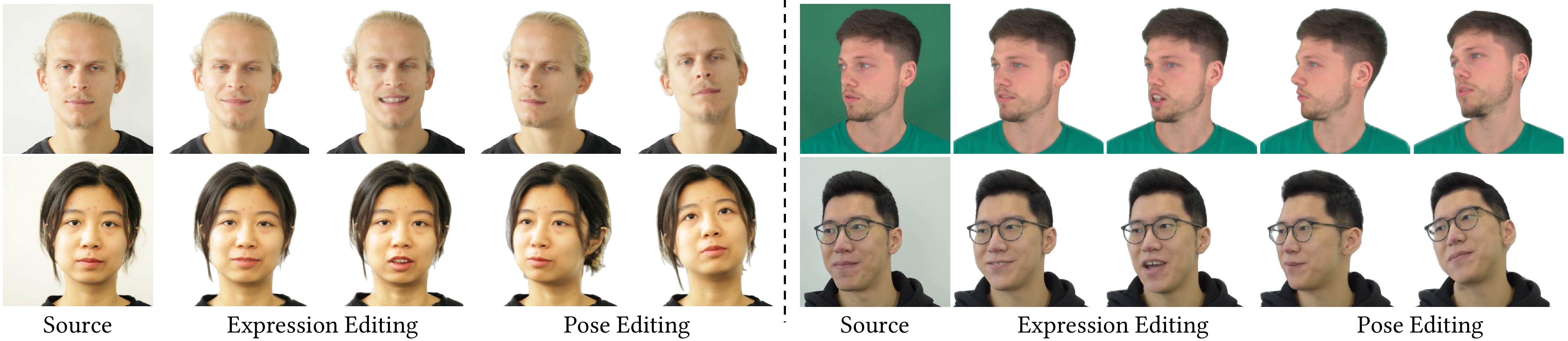}
  \vspace{-1.0mm}
  \caption{Expression and pose editing results. For each subject, the first column shows a source frame, followed by the editings of expressions (columns 2 and 3) and poses (columns 4 and 5).}
  \label{fig:face_editing}
\end{figure*}

Given a training portrait video, we execute self-reenactment driven by a testing video featuring the same subject and evaluate the results against SOTA methods. 
Specifically, we compare our approach with representative methods, 
including Bi-Layer~\cite{zakharov2020fast}, a one-shot image-based technique; NerFace~\cite{gafni2021dynamic}, a NeRF-based method; Neural Head Avatars (NHA)~\cite{grassal2022neural}, which employs an explicit 3D morphable mesh; IMavatar~\cite{zheng2022avatar}, which utilizes a neural implicit surface; PointAvatar~\cite{Zheng2023pointavatar}, a point clouds-based approach.

To evaluate image quality, we report $L_1$-distance, PSNR, SSIM, and LPIPS~\cite{zhang2018unreasonable} metrics. For expression accuracy of face reenactment, we measure the Keypoint Distance (KD) between the detected facial landmarks and the ground truth. We also report the average inference speed of all methods, measured on a desktop with an RTX 3090 GPU at $512 \times 512$ resolution. Our method's frame rate is measured in the interactive WebGL renderer, while other methods implemented on PyTorch~\cite{paszke2019pytorch} use a batch size of 1 during inference, simulating interactive face reenactment.

Tab.~\ref{tab:comparison} lists the quantitative metrics for self-reenactment results. Note that since NHA~\cite{grassal2022neural} does not generate the cloth part, it is excluded from general comparisons. In summary, our method not only effortlessly surpasses the maximum screen refresh rate for rendering speed but also achieves the highest rendering quality when compared to SOTA methods. Fig.~\ref{fig:comparison} illustrates the qualitative comparison of all methods for various subjects. The results indicate that our approach excels in capturing finer details, particularly in complex structures like hair and mouth, surpassing the performance of the other techniques.

We present visual quality metrics at each phase in Tab.~\ref{tab:comparasion_stage} to gain insights into how the baking and fine-tuning steps impact this aspect. Notably, after the second baking phase, a decrease in visual quality metrics is observed. This decline is due to errors stemming from the discretization of geometric and appearance attributes, such as the resolution involved in the marching cubes operation, the simplification of mesh faces, and texture seam issues arising from UV parameterization. However, these errors are rectified during the third fine-tuning phase, resulting in quality metrics that actually exceed those recorded after the initial training stage. Improvement in this stage is attributed to the elimination of low-frequency bias inherent in coordinate-based MLPs, and the direct optimization of extracted textures that enhance high-frequency textural details. The fine-tuning phase also eradicates texture seams observed in the baking stage and specifically fine-tunes the LPIPS metrics by rendering complete images rather than sampled rays.

\begin{figure*}[t!]
  \centering
  \includegraphics[width=\textwidth]{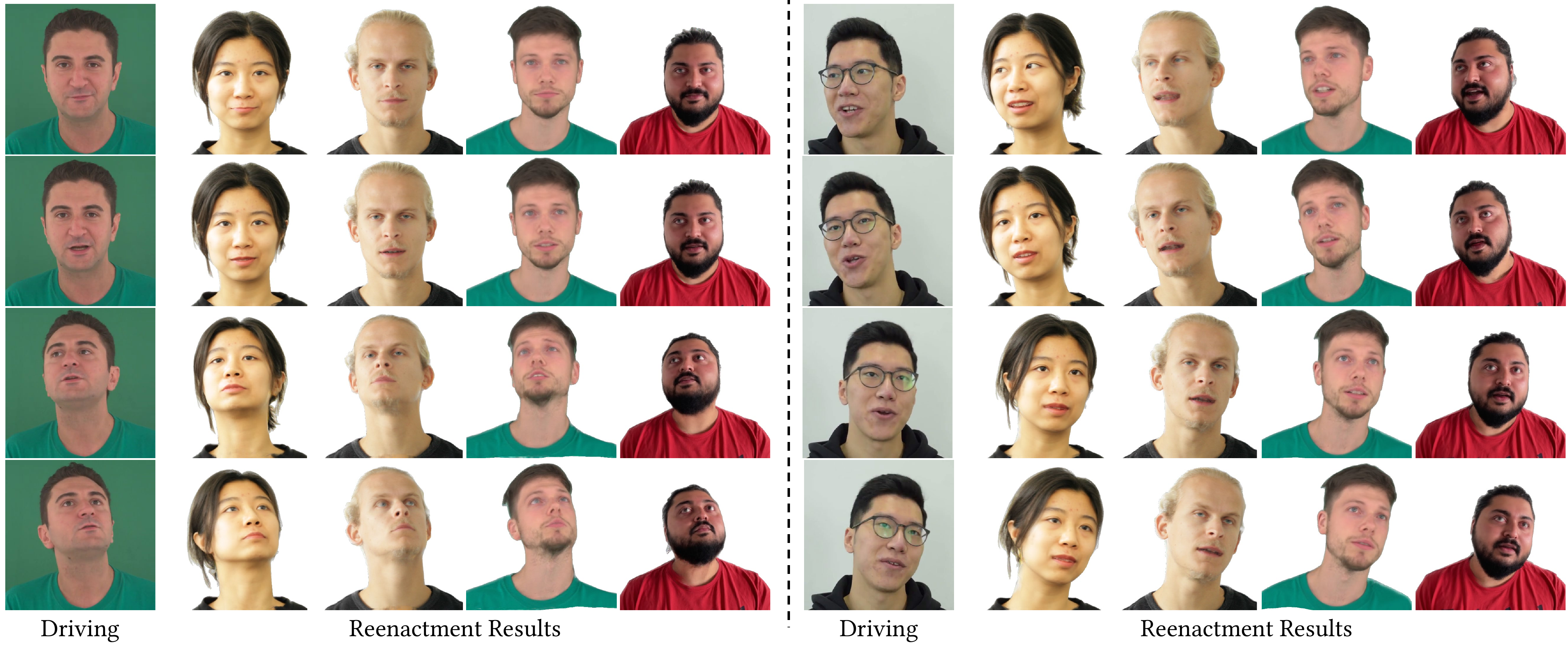}
  \vspace{-1.0mm}
  \caption{Cross-identity reenactment results. We use the expression and pose sequences extracted from a driving video to animate different subjects.}
  \label{fig:face_reenactment}
\end{figure*}

To evaluate the rendering performance of our method, we report the average frame rates (FPS) at different resolutions on commodity devices in Tab.~\ref{tab:speed}. Leveraging optimized triangle-based rasterization and a lightweight appearance decoder in pixel shaders, our approach delivers high-framerate rendering even on devices with limited computational power. To the best of our knowledge, \emph{BakedAvatar} is the first NeRF-based approach to achieve real-time animatable head avatars with SOTA quality on mobiles.

\begin{table}[!t]
  \caption{
  Rendering speed (FPS) of our method on commodity devices. The tested devices include a gaming laptop with an RTX 3060 Mobile GPU, a tablet (iPad Pro 2021), and a mobile phone (OnePlus 11). For the laptop, plugged and unplugged scenarios are tested. 
  The tablet and mobile devices had a maximum FPS cap of 60.
  }
  \resizebox{\columnwidth}{!}{%
  \label{tab:speed}
  \begin{tabular}{cccc}
    \toprule
    Resolution & Laptop (plugged/unplugged) & Tablet & Mobile\\
    \midrule
    $256\times256$ & $381$/$205$ & $60$ & $53$ \\
    $512\times512$ & $239$/$189$ & $59$ & $41$ \\
    $1024\times1024$ & $114$/$102$ & $27$ & $25$ \\
    $2048\times2048$ & $42$/$33$ & $9$ & $9$ \\
  \bottomrule
\end{tabular}
}
\end{table}

\subsection{Controlled Head Avatar Synthesis}

Since \emph{BakedAvatar} targets real-time rendering of reenactable head avatars, it is crucial to evaluate the synthesis quality of manually controlled novel expression, pose and view. Fig.~\ref{fig:view_synthesis} illustrates the results of novel view synthesis. We conduct editing by adding an offset to the FLAME expression and pose coefficients that represent novel facial expressions not covered during training. Fig.~\ref{fig:face_editing} demonstrates the expression and pose editing results for source frames, indicating that \emph{BakedAvatar} learns a generalized dynamic geometry and appearance representation of the head avatar. Additionally, we show cross-identity face reenactment results in Fig.~\ref{fig:face_reenactment}, where driving expression and pose come from different subjects. More results are provided in the supplementary video.

\begin{table}[!t]
  \caption{Ablation of rendering quality and speed.}
  \label{tab:ablation}
  \resizebox{1.0\columnwidth}{!}{%
  \begin{tabular}{p{4.7em}P{1.2em}ccccc}
    \toprule
    & & L1$\downarrow$ & PSNR$\uparrow$ & SSIM$\uparrow$ & LPIPS$\downarrow$ & FPS$\uparrow$\\
    \midrule
    \multirow{3}{*}{Mesh Layer}
    & $1$ & $0.0188$ & $25.47$ & $0.913$ & $0.0757$ & $781$ \\
    & $2$ & $0.0179$ & $25.88$ & $0.918$ & $0.0695$ & $435$ \\
    & $4$ & $0.0172$ & $26.34$ & $0.921$ & $0.0663$ & $223$ \\
    \midrule
    \multirow{4}{*}{Texture Basis}
    & $1$ & $0.0189$ & $25.28$ & $0.910$ & $0.0723$ & $573$ \\
    & $2$ & $0.0177$ & $25.93$ & $0.919$ & $0.0673$ & $220$ \\
    & $4$ & $0.0173$ & $26.19$ & $0.920$ & $0.0662$ & $183$ \\
    & $8$ & $0.0168$ & $26.59$ & $0.923$ & $0.0645$ & $156$ \\
    \midrule
    \multirow{3}{*}{Mesh Prec.}
    & $\times 1/2$ & $0.0173$ & $26.04$ & $0.921$ & $0.0680$ & $144$ \\
    & $\times 2$ & $0.0159$ & $27.00$ & $0.927$ & $0.0629$ & $89$ \\
    & $\times 4$ & $0.0158$ & $27.43$ & $0.925$ & $0.0648$ & $65$ \\
    \midrule
    \multirow{2}{*}{Texture Res.}
    & $256^2$ & $0.0177$ & $26.05$ & $0.919$ & $0.0687$ & $134$ \\
    & $512^2$ & $0.0167$ & $26.54$ & $0.923$ & $0.0644$ & $127$ \\
    \midrule
    \multicolumn{2}{l}{w/o fine-tuning}
    & $0.0185$ & $26.16$ & $0.917$ & $0.0937$ & $-$ \\
    \multicolumn{2}{l}{w/o cam\&pose opt.}
    & $0.0185$ & $25.80$ & $0.914$ & $0.0720$ & $-$ \\
    \multicolumn{2}{l}{w/o VGG loss}
    & $0.0167$ & $26.53$ & $0.924$ & $0.0762$ & $-$ \\
    \midrule
    \multicolumn{2}{l}{Ours (full)}
    & $0.0163$ & $27.13$ & $0.925$ & $0.0643$ & $114$ \\
    \bottomrule
  \end{tabular}
  }
\end{table}

\begin{figure*}[t!]
  \centering
  \includegraphics[width=\textwidth]{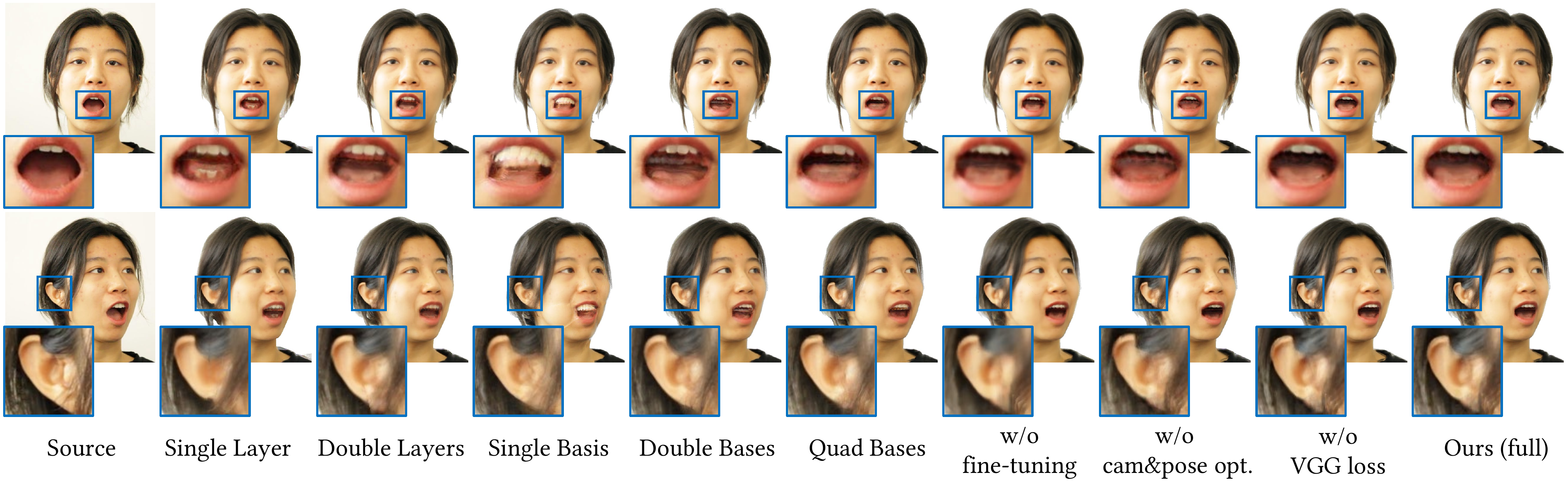}
  \vspace{-1.0mm}
  \caption{Ablation study results. Given a source frame (column 1), we visually assess the effect of mesh layers (columns 2 and 3), radiance bases (columns 4 to 6), fine-tuning (column 7), camera and pose optimization (column 8), and perceptual loss (column 9) on self-reenactment results.}
  \label{fig:ablation}
\end{figure*}

\subsection{Ablation Studies}

We report the rendering quality and speed of our full model and other design choices. FPS is measured on an RTX 3060 Mobile GPU.

\subsubsection{Mesh Layers}

We compare the quality and speed with various numbers of mesh layers. For a single mesh layer, our method is the same as traditional morphable mesh representation, with an expression- and pose-conditioned neural texture. The results, presented in Tab.~\ref{tab:ablation} and Fig.~\ref{fig:ablation}, indicate that increasing the number of mesh layers significantly improves quality. Nevertheless, this improvement is accompanied by slower rendering times. 
Our multi-layer meshes effectively approximate volumetric effects and achieve better rendering quality for complex structures.

\subsubsection{Radiance Bases}

As our expression- and pose-dependent appearance decouples dynamic appearance as multiple radiance texture bases, we compare the impact on rendering quality with various numbers of bases. When the number of bases is set to $1$, the appearance is identical to a fully static diffuse color representation. As shown in Tab.~\ref{tab:ablation} and Fig.~\ref{fig:ablation}, the single basis fails to represent dynamic appearance caused by expression and pose changes, and increasing the number of texture bases gradually improves quality.

\subsubsection{Mesh and Texture Complexity} \label{mesh_precision}

We further analyze how the precision of extracted meshes and textures affects the final rendering quality and speed. The results from different configurations of mesh (expressed as multiples of the target triangle count) and texture extracting setups are listed in Tab.~\ref{tab:ablation}.
As we encourage smooth geometry in Sec.~\ref{canonical_manifold}, the precision of the mesh extraction process does not significantly affect the synthesis quality of our representation. Thus we can utilize a simple mesh with a moderate number of triangles so as to be rendered with lower computation cost.

\begin{figure}[!t]
  \includegraphics[width=\columnwidth]{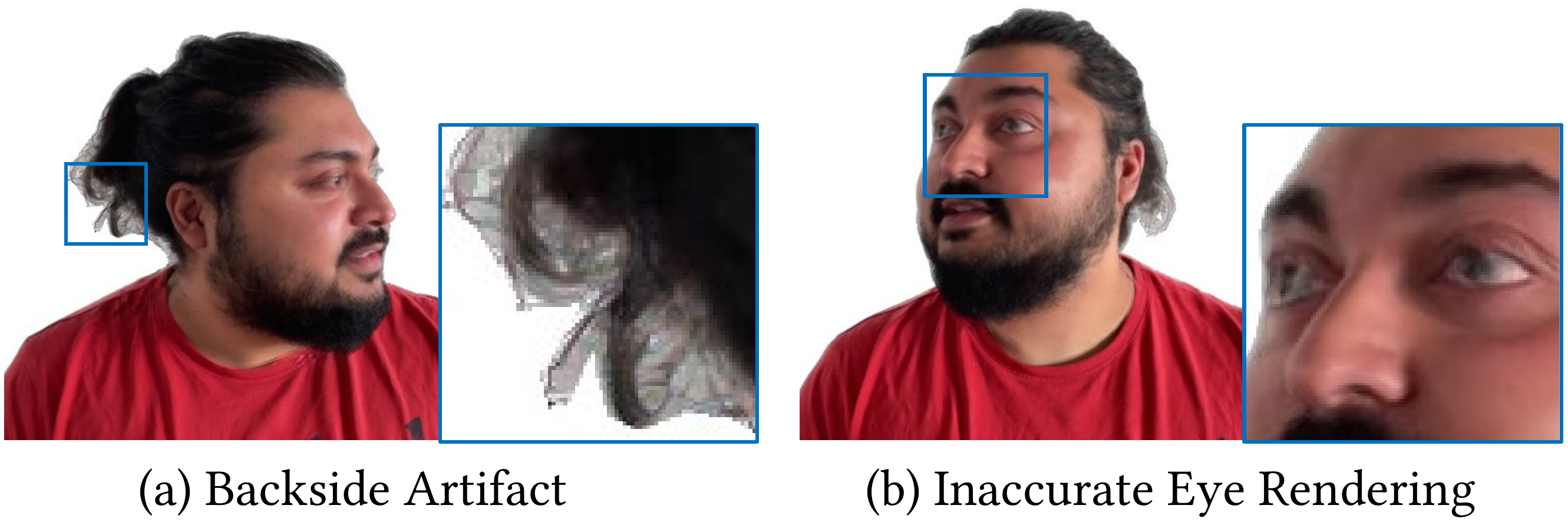}
  \vspace{-1.0mm}
  \caption{Failure cases. (a) Artifact occurs when rendering the backside of the head avatar. (b) \emph{BakedAvatar} does not specifically model eyeballs, which may lead to inaccurate eye rendering.}
  \label{fig:failure}
\end{figure}

\begin{figure}[!t]
  \includegraphics[width=\columnwidth]{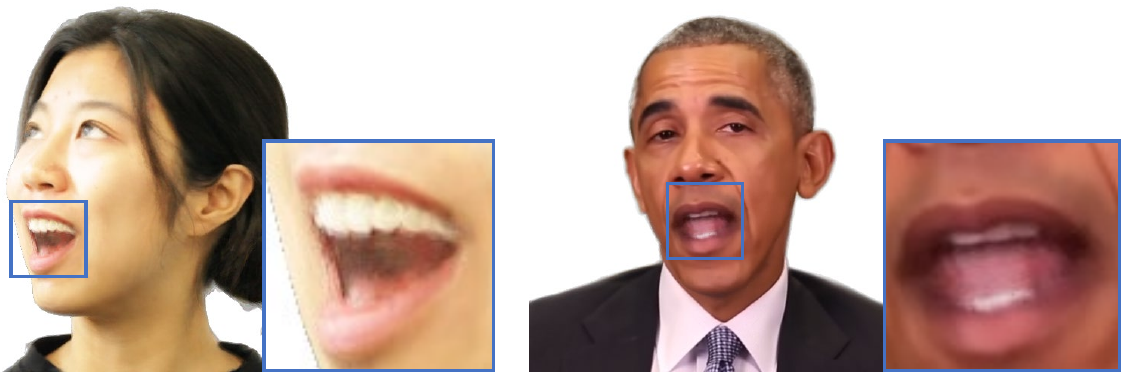}
  \vspace{-1.0mm}
  \caption{
  Details of the mouth region with extrapolated expressions that are not well covered by the training set. Although such expression extrapolation is feasible, the visual quality might be inferior due to the difficulty of accurate reconstruction of geometry and deformation around the mouth interior and teeth.
  The right figure is from \emph{Obama\textcopyright The Obama White House}.
  }
  \label{fig:mouth_region}
\end{figure}

\subsubsection{Fine-tuning}

We assess the quality enhancement of our comprehensive model compared to alternative design choices during the fine-tuning phase, as presented in Table~\ref{tab:ablation} and Figure~\ref{fig:ablation}. Specifically, we compare our complete method with our approach excluding the fine-tuning phase, without camera and pose tuning, and without the incorporation of perceptual loss. The absence of fine-tuning results in the omission of high-frequency appearance details. Furthermore, the inclusion of camera and pose tuning, as well as perceptual loss, contributes to improved learning of intricate structures.

\section{Conclusion}

We present \emph{BakedAvatar}, a new representation for synthesizing person-specific 4D head avatars. Our key idea is to bake the learned deformation, geometry, and appearance neural fields into layered meshes and multiple texture bases, combining with an efficient pose-, expression- and view-dependent appearance representation that can be evaluated with lightweight computation. We demonstrate that our method is able to achieve SOTA-quality synthesis results while running at real-time frame rates with the rasterization pipeline, enabling interactive applications such as facial reenactment and expression editing, even on mobile devices with limited computation resources.

\paragraph{Limitations and Future work}
The capability of \emph{BakedAvatar} is limited by the number of mesh layers and the detail quality of mesh textures, and may not be able to achieve full volumetric and refraction effects faithfully, such as thin hair strands and distortion of eyeglasses. Our methods can extrapolate to unseen head poses and facial expressions that are not covered by the training data, nevertheless, the synthesis outcome for extreme views or expressions may be of inferior quality. Fig.~\ref{fig:failure} shows failure examples. In particular, visual artifacts of complex novel pose and expression around the mouth region might occur, as demonstrated in Fig.~\ref{fig:mouth_region}. This is possibly caused by the mismatch between the learned FLAME weights of extracted meshes and the ground truth, as it is challenging to learn deformation accurately due to appearance and geometry complexity in the mouth region. This can be addressed with a more accurate reconstruction of deformation and a non-linear representation of the dynamic appearance, which we leave as future work. 

During the dataset pre-processing steps, we adopt a FLAME tracking pipeline from \cite{zheng2022avatar}, which initially estimates FLAME shape and expression parameters with DECA~\cite{DECA:Siggraph2021}, and then optimizes them with detected facial keypoints, leveraging loss functions based on assumptions of temporal sequences. While the derived FLAME parameters are sufficient for implicit field reconstruction, flickering artifacts can be observed for specific identities (Fig.~\ref{fig:comparison}, Row 1-2) in the facial reenactment task. This issue is likely to be attributed to the temporal inconsistencies in the detected facial keypoints of consecutive frames. Our attempts indicate that slightly smoother motion can be obtained with a better facial landmark detector~\cite{bazarevsky2019blazeface} and higher weights on temporal consistency loss. Still, better results might be achieved with a photometric tracker, or temporal consistent optimization of the FLAME parameter sequences during training. However, as the focus of this paper is not on facial tracking techniques, we leave the improvements as future work.

Our method can be further extended in several directions.
To facilitate relighting applications, the appearance representation can be further divided into distinct elements like environment lighting and albedo/specular maps, integrating these with a physics-based lighting model. To enhance the precision and realism of iris movements, incorporating specific modeling for eyeballs could be beneficial. Expanding our technique to generate head avatars that are person-agnostic could also be an interesting direction to explore.

\begin{acks}
The authors thank the anonymous reviewers for their insightful feedback and also appreciate the consent given by authors from other publications to use video datasets of the featured individuals. This work was supported by the National Natural Science Foundation of China (Project Number: 61932003) and the Fundamental Research Funds for the Central Universities. 
\end{acks}

\bibliographystyle{ACM-Reference-Format}
\bibliography{paper}

\appendix

\section{Appendix}

\subsection{Network Architecture} \label{network_arch}

\begin{figure}[h]
  \includegraphics[width=\columnwidth]{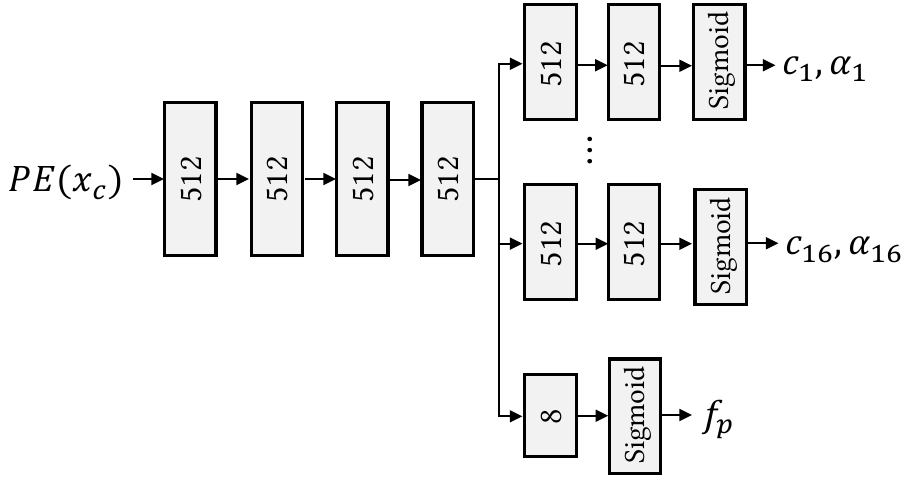}
  \caption{Network structure of the canonical radiance field. There are multiple heads for outputting $n_T$ radiance bases ($n_T = 16$ in the main paper), and a head for outputting the shared position feature $f_p$.}
  \label{fig:network}
\end{figure}

In the training stage, we first learn all fields as coordinate-based implicit MLPs. Specifically, we use a $6$ layer MLP with $256$ hidden neurons for the manifold network, and a $3$ layer MLP with $128$ hidden neurons for the deformer network, both with Softplus activation. Since the radiance network outputs multiple radiance bases, we design it as a trunk of a $4$ layer MLP, a head that outputs the shared position feature $f_p$, and several separate heads of a $2$ layer MLP that outputs radiance $c_i$ and occupancy $\alpha_i$ of each texture basis, all with $512$ hidden neurons and ReLU activation (except for the last layer), as illustrated in Fig.~\ref{fig:network}. Positional encoding with frequency numbers $5, 12$ are used for the manifold and radiance network respectively, while the deformer network does not use positional encoding. 

For the appearance decoder, we use a $3$-layer MLP of $64$ hidden neurons with PReLU activation for the global MLP and a $2$-layer MLP with $16$ hidden neurons for the spatial MLP. After the training and fine-tuning stage, the global MLP weights of the appearance decoder are exported, together with the meshes and textures.

\subsection{Training of the Forward Deformer}  \label{forward_deformer}

In Section~\ref{deformer}, we learn the FLAME blendshapes and linear blend skinning weights in canonical space. As the FLAME template has a closed mouth state, causing a sharp transition of LBS weights that can not be learned well by MLP, we adopt the approach from PointAvatar~\cite{Zheng2023pointavatar} by introducing a custom canonical space with a half-opened mouth state. Specifically, we first warp the points in the custom canonical space to the original FLAME space, then use FLAME expression and pose coefficients to warp them to the deformed space with learned blendshapes and LBS weights. When exporting the extracted multi-layer meshes at the baking stage in Section~\ref{stage2} of the main paper, we transform all vertex positions and normals from canonical space to the original FLAME template space, removing the first warping procedure at inference time.

\subsection{Implementation details of real-time rendering}

\begin{table}
  \caption{Total polygon count and uncompressed storage size (MiB) under various mesh precision and texture resolution settings.}
  \label{tab:mesh}
  \begin{tabular}{lccP{2.4em}P{2.4em}P{2.4em}}
    \toprule
    \multirow{2}{*}{Mesh Precision} & \multirow{2}{*}{Vertices} & \multirow{2}{*}{Triangles} & \multicolumn{3}{c}{Storage under tex. size} \\
    & & & $256^2$ & $512^2$ & $1024^2$ \\
    \midrule
    $\times 1/2$ & $13544$ & $19017$ & $41$ & $84$ & $248$ \\
    $\times 1$ (Ours) & $28882$ & $44298$ & $58$ & $107$ & $266$ \\
    $\times 2$ & $61401$ & $100492$ & $90$ & $135$ & $302$ \\
    $\times 4$ & $128540$ & $220570$ & $161$ & $206$ & $374$ \\
    \bottomrule
  \end{tabular}
\end{table}

In the baking stage outlined in Section~\ref{stage2}, we employ a mesh simplification scheme to reduce the complexity of the extracted mesh to a desired number of triangles. To assess the complexity of the resulting meshes, we provide the total count of vertices and triangles for the extracted layered meshes, along with their storage requirements at different mesh precision and texture resolutions, as shown in Table \ref{tab:mesh}.

During the texture export process, we utilize quantization-aware training to address quantization errors that may arise when converting continuous color values into 8-bit integers, by clamping and rounding values during the forward pass while retaining their floating-point gradient in the backward pass. Upon completion of the fine-tuning process, we export two expanded RGBA textures: the position texture and multiple radiance texture bases, by tiling the multi-channel image as a grid of images on a large texture atlas, which allows better compression with PNG and avoiding exceeding the texture unit limit in the WebGL implementation.

To ensure the correct rendering order for the alpha composition result, as described in Section~\ref{RTR} of the main paper, we draw the multi-layer meshes from the inner layers to the outer layers. However, the rasterization order of triangles within the same mesh layer is not deterministic, which may lead to incorrect alpha composition results. To deal with the ordering problem of transparent triangles in the same layer, each mesh layer is rendered in two passes: the first pass acquires the depth of forefront triangles by only writing to the depth buffer, and the second pass draws pixels with depth less or equal than the current value in the depth buffer using depth test. This approach effectively occludes each mesh layer with minimal additional computational overhead.

\subsection{Details of Performance Measurement} \label{perfermance_measurement}

The performance of our representation is measured with the WebGL2 interactive renderer using browsers under each device. For the desktop and laptop computers, we close all other applications and disable the frame rates limit and vertical synchronization in Chrome by setting:

\begin{quote}
\texttt{----disable-gpu-vsync ----disable-frame-rate-limit}
\end{quote}

We measure the average frame rates of 6 avatar subjects when rendering a segment of the facial reenactment sequence. We set the camera field-of-view to $14^{\circ}$ to ensure the head avatar roughly covers the entire rendering canvas. When evaluating baseline methods implemented in the PyTorch framework, we measured their average inference time on a desktop computer equipped with an NVIDIA RTX 3090 GPU. We use a batch size of $1$ with gradient computation disabled to simulate the facial reenactment application with real-time user inputs, and exclude all irrelevant Python operations. 

To compare the energy efficiency of different methods, we provide the performance-to-watt ratios for our method and other approaches in Table \ref{tab:perfermance}. Our representation demonstrates higher energy efficiency compared to alternative head avatar synthesis methods. Consequently, it enables real-time head avatar synthesis on devices with highly limited computational resources, such as tablets and mobile phones. The primary factor contributing to the reduced computational cost of our method is the baked radiance representation, which eliminates the need for evaluating large neural networks during inference. Additionally, our approach employs rasterization-based rendering instead of the more computationally expensive ray-marching procedure.
While Neural Head Avatar~\cite{grassal2022neural} and PointAvatar~\cite{grassal2022neural} also utilize rasterization-based rendering, their runtime performance remains constrained by the size of their neural networks.

\begin{table}[!h]
  \caption{Energy Efficiency Comparison. FPS is measured under $512 \times 512$ resolution.
  }
  \label{tab:perfermance}
  \begin{tabular}{lccc}
    \toprule
    Method & Watts$\downarrow$ & FPS$\uparrow$ & FPS/W$\uparrow$\\
    \midrule
    BiLayer~\shortcite{zakharov2020fast} & $350$ & $3.39$ & $0.00968$\\
    NerFace~\shortcite{gafni2021dynamic} & $350$ & $0.15$ & $0.00042$\\
    IMavatar~\shortcite{zheng2022avatar} & $350$ & $0.02$ & $0.00005$\\
    PointAvatar~\shortcite{Zheng2023pointavatar} & $350$ & $8.53$ & $0.02437$\\
    NHA~\shortcite{grassal2022neural} & $350$ & $12.30$ & $0.03514$\\
    Ours (Desktop) & $350$ & $804$ & $2.29714$\\
    Ours (Laptop) & $105$ & $239$ & $2.27619$\\
    Ours (Tablet) & $17$ & $59$ & $3.47058$\\
    Ours (Mobile) & $13$ & $41$ & $3.15384$\\
  \bottomrule
\end{tabular} 
\end{table}

\subsection{Ethical Considerations}

In the research presented within this paper, we introduce novel techniques capable of reconstructing high-fidelity dynamic 3D avatars from monocular videos and performing expression transfer in real-time across various devices, including mobile platforms. This method's capacity to synthesize a realistic rendering of individuals, with detailed control over expression and head pose, is not devoid of significant ethical considerations. Specifically, these techniques harbor the potential for malevolent utilization, such as fabricating authentic-seeming portrayals of individuals within contexts they have never engaged in, consequently misleading viewers and possibly resulting in defamation or harm to the depicted persons. Additionally, the ability of these techniques to extrapolate from existing images can engender potential privacy concerns, by enabling the depiction of individuals in unauthorized or invasive scenarios. While the primary aim of this technology serves to foster positive applications, such as enhancing connections through mixed-reality video conferencing or the development of virtual human guides, we cannot overlook its potential for misuse. The eradication of technologies akin to deep-fakes~\cite{nguyen2022deep} may be beyond our reach, yet we aspire that our contributions will augment the understanding of human avatar synthesis, thereby informing the development of countermeasures against any unscrupulous exploitation of this technology.

\end{document}